# Detachment of adhesive normal contact between a rigid circular flat punch and a viscoelastic half-space

Emanuel Willert, e.willert@tu-berlin.de
Valentin L. Popov, v.popov@tu-berlin.de

Technische Universität Berlin, 10623 Berlin, Germany

**Abstract.** We propose an approach to describe the propagation of a crack (or boundary of an adhesive contact) in a viscoelastic material which is only based on the consideration of the rheology of the material without the introduction of any additional dependency of the separation energy on the velocity of crack propagation. The suggested idea is illustrated with an example of kinetics of detachment of a flat-ended indenter from a viscoelastic medium. It is shown that under the given assumptions the crack propagation is accelerating until the critical configuration is reached and the contact detaches instantaneously. The suggested criterion can be basically applied to arbitrary shapes and arbitrary loading histories.

## 1 Introduction

Adhesion of viscoelastic bodies is a phenomenon that has been studied for several years, but which has not yet been understood in its principles [1], [2], [3], [4], [5]. There are no universal and recognized approaches as the principle of energy balance in the case of elastic contacts. Accordingly, there are no physically sound, and certainly no numerically efficient, simulation methods in this area.

The aim of this manuscript is to investigate a new approach to the modeling and computation of such contacts. The physical basis for this approach is the idea, which has emerged from many experimental studies, that the detachment of the adhesive contact is associated with a strong elongation of the interface molecules. For the detachment on the microscopic scale, two scenarios are conceivable:

- Peel-off occurs when the stress near the crack tip (the boundary of the adhesive contact represents an "external" crack) reaches a critical value (at the molecular distance from the crack tip, the stress must meet the theoretical strength limit to separate molecules from each other).

- Peel-off occurs when molecules close to the crack tip have been deformed to a critical length.

The first criterion we call "stress criterion", the second "deformation criterion". It is clear that for linear-elastic materials both criteria are equivalent. For viscoelastic media, however, this is not the case. It was experimentally established that the predictions of the stress criterion do not correspond to the empirical behavior of viscoelastic materials. So we will focus on the deformation criterion.

## 2 Problem formulation

We want to study the following contact problem: a rigid circular flat punch with radius $a_0$ is brought into contact with an incompressible, linearly viscoelastic medium with the shear creep function $J(t)$, giving the deformation response to a unit of instantaneously applied shear stress. In the contact there shall be adhesive interaction with the effective work of separation per unit area, $\Delta\gamma$.

The punch is pulled upwards extremely slowly, so that the viscoelastic medium will behave elastically with the static compliance $J_\infty$. If the vertical displacement of the flat punch reaches the critical elastic value given by Kendall [6],

$$d = \sqrt{\frac{2\pi\Delta\gamma}{E^*_\infty}} a_0 = \sqrt{\frac{\pi\Delta\gamma}{2} a_0 J_\infty}, \qquad (1)$$

an external crack will propagate inside from the edge of the punch and the contact will start to detach. If the medium were elastic, the crack would propagate infinitely fast and the contact would detach instantly. However, due to the viscoelastic properties the crack will, at least for a certain period of time, propagate with finite velocity, even in the quasi-static limit. When the contact starts to detach, we would like to fix the punch displacement at the value given in Eq. (1).

## 3 Relaxation of a single rheological element

The frictionless normal contact problem between an axisymmetric rigid indenter and a linearly viscoelastic half-space can be exactly mapped onto an equivalent contact between a properly defined rigid plain profile and a one-dimensional foundation of independent linearly viscoelastic elements [7]. We thus consider first the deformation process of a single rheological element and later perform the transformation back to the real space. The creep function of the rheological elements is given by the function $J$ of the original viscoelastic half-space. Thus, if the element is detached (and therefore free of forces) from the static displacement $d$, the resulting time-dependent displacement will be

$$w(t) = d\left[1 - \frac{J(t)}{J_\infty}\right]. \qquad (2)$$

Note, that if the viscoelastic medium has a non-vanishing instantaneous compliance, i.e. a finite glass modulus, there will be an instantaneous jump before the relaxation of

$$\Delta w = \frac{J(t=0)}{J_\infty} d := \delta d. \qquad (3)$$

For very small times $t$, Eq. (2) can be linearized to give

$$w(t) \approx d\left[1 - \delta - \frac{t}{\tau}\right], \quad \tau := \frac{J_\infty}{\dot{J}(t=0)}, \qquad (4)$$

where dot denotes the time derivative.

## 4 A viscoelastic detachment criterion

The physical idea of the criterion of detachment for one element of the viscoelastic foundation is based on the deformation criterion described in the introductory section: the external crack propagates, if in the actual three-dimensional system, the elongation in a molecular distance $b$ from the edge of contact with radius $a$,

$$f(r = a + b) := d - w^{3D}(r = a + b), \qquad (5)$$

reaches a critical value. The displacements in the equivalent one-dimensional system are, close to the edge of contact, given by

$$w^{1D}(x) = \begin{cases} d, & \text{for } |x| \le a, \\ \left(1 - \delta - \dfrac{x-a}{\tau c(a)}\right) d, & \text{for } |x| > a, \end{cases} \qquad (6)$$

whereas $c(a)$ denotes the current velocity of crack propagation. The one-dimensional and three-dimensional displacements are coupled via the Abel-transform [7]

$$w^{3D}(r) = \frac{2}{\pi} \int_0^r \frac{w^{1D}(x)}{\sqrt{r^2 - x^2}} \, dx. \qquad (7)$$

If we introduce the small parameter

$$\varepsilon := \frac{b}{a}, \qquad (8)$$

perform the integration, develop the result in a Taylor series with the argument $\sqrt{2\varepsilon}$ and only consider the first non-vanishing order for both the components (one for the instantaneous jump and one for the relaxation outside the contact), we arrive at the asymptotic solution

$$f(\varepsilon) \approx \frac{2d}{\pi}\left[\delta\sqrt{2\varepsilon} + \frac{2b}{3\tau c(a)}\sqrt{2\varepsilon}\right]. \qquad (9)$$

Hence, the detachment criterion will be

$$d = \frac{C\sqrt{a}}{\delta + 2b/(3\tau c(a))}, \qquad (10)$$

whereas the constant $C$ follows from the elastic solution

$$C = \sqrt{\frac{\pi \Delta \gamma J_\infty}{2}} = \frac{d}{\sqrt{a_0}}. \qquad (11)$$

## 5  Propagation of the viscoelastic crack

The detachment criterion (10) can be written in the form

$$\sqrt{a_0} = \frac{\sqrt{a}}{\delta + c_0/c(a)}, \quad c_0 := \frac{2b}{3\tau}. \qquad (12)$$

Under the circumstances considered here, the velocity of crack propagation is always a positive quantity, because contact is never re-established anywhere. Therefore, the last relation can only be fulfilled as long as

$$\frac{a}{a_0} > \delta^2. \qquad (13)$$

If the contact radius reaches this critical value, the crack starts to propagate infinitely fast and the contact is instantaneously detached. The velocity of crack propagation is given by

$$c(a) = -\dot{a}, \qquad (14)$$

so Eq. (12) gives a differential equation for the development of the contact radius. Introducing the normalized variables

$$\alpha := \frac{a}{a_0}, \quad \tau := \frac{t}{t_0}, \quad t_0 := \frac{a_0}{c_0}, \tag{15}$$

this differential equation reads

$$\frac{d\alpha}{d\tau} = \frac{1}{\delta - \sqrt{\alpha}}, \quad \text{for } \delta < \sqrt{\alpha}, \tag{16}$$

with the implicit solution (the initial condition is $\alpha(\tau=0)=1$)

$$\tau = -\delta(1-\alpha) + \frac{2}{3}(1-\alpha^{3/2}). \tag{17}$$

The duration of the detachment process is given by setting $\alpha = \delta^2$, hence

$$T = -\delta(1-\delta^2) + \frac{2}{3}(1-\delta^3). \tag{18}$$

The development of the contact radius as a function of time is shown in Fig. 1. Obviously the crack propagation is accelerating until the critical configuration is reached and the contact detaches instantaneously.

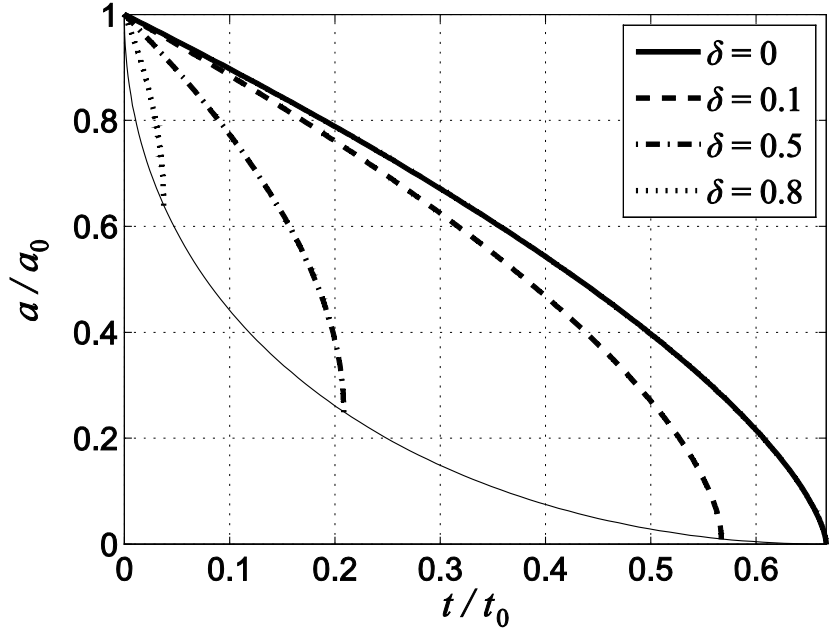

**Fig. 1** Development of the contact radius as a function of time for the detachment of an adhesive normal contact between a rigid circular flat punch and a viscoelastic half-space for different values of the instantaneous jump $\delta$. The thin line denotes the series of critical configurations, after which the crack starts propagating infinitely fast.

## 6  Implementation in the Boundary Element Method

The presented idea can be easily implemented in the framework of the Method of Boundary Elements by using a mesh-dependent detachment criterion very much similar to how it has been done in [8] in the case of purely elastic contacts. Let us consider a simulation area discretized in square elements with the size $\Delta x = \Delta y = \Delta$. In each time step, the outputs of the program are stresses inside the contact area and the gap outside the contact. We consider an element outside the contact area but directly adjacent to the boundary of the contact. If the gap in this element is larger than the critical one given by the condition

$$u_{z,c} = \frac{C\sqrt{\Delta}}{\delta + c_0/c}, \quad (19)$$

then the next adjacent element inside the contact area detaches. Here $c = \Delta/(n\Delta t)$ is the current "assumed propagation velocity" of the crack, $\Delta t$ is the time step and $n$ is the number of time steps since the detachment of the element. The constants $C$ and $c_0$ are basically given by (11) and (12):

$$C = \sqrt{\frac{\pi \Delta \gamma J_\infty}{2}}, \quad c_0 := \frac{2b}{3\tau}. \quad (20)$$

However, some universal dimensionless correction factors may be necessary as discussed in [9] in the case of gradient materials.

# 7 Conclusions

We proposed a deformation-based detachment criterion for the adhesive normal contact of viscoelastic materials. The detachment process from a rigid circular flat punch has been studied in detail. We find that the adhesive crack, after a stage of propagation with finite velocity, depending on the viscoelastic rheology reaches a critical state with following instant detachment, likewise the elastic case. Although the detachment criterion was derived from considerations on axisymmetric contacts, it can be generalized and implemented as a mesh-dependent criterion in Boundary Element Simulations of viscoelastic adhesive contacts without restrictions regarding the contact geometry.